\documentclass[pra, twocolumn]{revtex4}
\usepackage{epsfig}
\def\beq{\begin{equation}}
\def\eeq{\end{equation}}
\def\bea{\begin{eqnarray}}
\def\eea{\end{eqnarray}}
\def\ba{\begin{array}}
\def\ea{\end{array}}
\def\bay{\begin{array}}
\def\eay{\end{array}}

\def\bb{{\bf b}}
\def\be{{\bf e}}
\def\bk{{\bf k}}

\def\pp{{\bf p}}
\def\bp{{\bf p}}
\def\bq{{\bf q}}

\def\bx{{\bf x}}
\def\by{{\bf y}}
\def\bz{{\bf z}}

\def\hbk{\hat{\bk}}
\def\hbx{\hat{\bx}}
\def\hby{\hat{\by}}
\def\hbz{\hat{\bz}}

\def\0{\otimes}
\def\1{\mbox{\small1\hskip-0.35em\normalsize1}}
\def\6{\langle }
\def\9{\rangle }
\def\half{\mbox{$1\over2$}}
\def\tr{{\rm tr}}

\def\bep{\mbox{\boldmath $\epsilon$}}
\def\bal{\mbox{\boldmath $\alpha$}}

\begin{document}

\renewcommand{\thefootnote}{\fnsymbol{footnote}}

\title{Quantum information and special relativity}
\author{Asher Peres and Daniel R. Terno}
\address{
 Department of Physics,
Technion---Israel Institute of Technology,
 32000 Haifa, Israel}
%\email{terno@physics.technion.ac.il.}
%\date{}

\begin{abstract}
Relativistic effects affect nearly all notions of quantum
information theory. The vacuum behaves as a noisy channel, even if
the detectors are perfect.  The standard definition of a reduced
density matrix fails for photon polarization because the
transversality condition behaves like a superselection rule. We
can however define an effective reduced density matrix which
corresponds to a restricted class of positive operator-valued
measures. There are no pure photon qubits, and no exactly
orthogonal qubit states. Reduced density matrices for the spin of
massive particles are well-defined, but  are not covariant under
Lorentz transformations. The spin entropy is not a relativistic
scalar and has no invariant meaning. The distinguishability of
quantum signals  and their entanglement depend on the relative
motion of observers.
\end{abstract}
\maketitle

\section{Introduction}
The relationship between information and physics has been an
intriguing problem for many years \cite{inph}. It took a new twist
with the emergence of quantum information theory whose paradigm
 ``information is physical"
\cite{lanben} added a new point of view to old questions. Quantum information
 theory usually involves only  a
nonrelativistic quantum mechanics. However, both for the sake of
logical completeness and in order to derive physical bounds  on
 information transfer, its processing, and the errors involved, a
full relativistic treatment is required. Additional motivation to
relativistic extensions of quantum information theory comes from
quantum cosmology. Quantum field theory in curved spacetime, and
black hole physics in particular, present  challenges that
everybody who upholds the principle that ``information is
physical" should respond to. Techniques of quantum information
theory can be applied to  field theory and black hole physics
\cite{pt03b}. In this presentation we describe some of the new
features of quantum information theory when the effects of special
relativity are taken into account.

The concept of  reduced density matrix is fundamental for quantum
information. Its properties are significantly modified when we
deal with relativistic effects. Let us use Latin indices for the
description of a subsystem which is excluded from our description,
and Greek indices for the subsystem that we can actually describe.
The components of a state vector would thus be written $V_{m\mu}$
and those of a density matrix $\rho_{m\mu,n\nu}$. The reduced
density matrix of the system of interest is given by
\beq
\tau_{\mu\nu}=\sum_{m,n} \rho_{m\mu,n\nu}. \label{tau}
\eeq
Even if $\rho$ is a pure state (a matrix of rank one), $\tau$ is
in general a mixed state. Its {\it entropy\/} is defined as
\beq
S=-\tr(\tau\log\tau).
\eeq

An important consequence of relativity is that there is a
hierarchy of dynamical variables: {\it primary variables\/} have
relativistic transformation laws that depend only on the Lorentz
transformation matrix $\Lambda$ that acts on the spacetime
coordinates. For example, momentum components are primary
variables. On the other hand, {\it secondary variables\/} such as
spin and polarization have transformation laws that depend not
only on $\Lambda$, but also on the momentum of the particle. As a
consequence, the reduced density matrix for secondary variables,
which may be well defined in any coordinate system, has no
transformation law relating its components in different Lorentz
frames. A simple example will be given below.

Moreover, an unambiguous definition of the reduced density matrix
by means of Eq.~(\ref{tau}) is possible only if the secondary
degrees of freedom are unconstrained. For gauge field theories,
that equation may be meaningless if it conflicts with constraints
imposed on the physical states \cite{bgkp02, pt03a}. In the
absence of a general prescription, a case-by-case treatment is
required. A particular construction, valid with respect to a
certain class of tests, is given in Sec.~\ref{photons}.  A general
way of defining reduced density matrices for physical states in
gauge theories is an open problem.

The next two sections present detailed calculations and explore
implications for distinguishability of quantum states and their
entanglement. In the last section we show that a description of
quantum channels by means of completely positive maps \cite{
pt03b, book} is only an approximation.

\section{Massive particles}

We first consider the relativistic properties of the spin entropy
for a single, free particle of spin~\half\ and mass $m>0$. For
massive particles a reduced density matrix is well-defined, but it
has no invariant meaning \cite{pst}. The reason is that under a
Lorentz boost, the spin undergoes a Wigner rotation
\cite{we:b,hal,bog}  whose direction and magnitude depend on the
momentum of the particle. Even if the initial state is a direct
product of a function of momentum and a function of spin, the
transformed state is not a direct product. Spin and momentum
become entangled.

The quantum state of a spin-\half\ particle can be written, in the
momentum representation, as a two-component spinor,
\beq
\psi(\pp)={a_1(\pp)\choose a_2(\pp)}, \label{psi0}
\eeq
where the amplitudes $a_r$ satisfy $\sum_r\int|a_r(\pp)|^2
d\pp=1$. The normalization of these amplitudes is a matter of
convenience, depending on whether we prefer to include a factor
\mbox{$p_0=(m^2+\pp^2)^{1/2}$} in it, or to have such factors in
the transformation law (\ref{tf}) below.  Here we shall use the
second alternative, because it is closer to the nonrelativistic
notation which appears in the usual definition of entropy.

We emphasize that we consider normalizable states, in the momentum
representation, not momentum eigenstates as usual in textbooks on
particle physics. The latter are chiefly concerned with the
computation of $\6\mbox{in}|\mbox{out}\9$ matrix elements needed
to obtain cross sections and other asymptotic properties. However,
in general a particle has no definite momentum. For example, if an
electron is elastically scattered by some target, the electron
state after the scattering is a superposition that involves
momenta in all directions. In that case, it still is formally
possible to ask, in any Lorentz frame, what is the value of a spin
component in a given direction (this is a legitimate Hermitian
operator).

Let us define a reduced density matrix, $\tau=\int
d\pp\,\psi(\pp)\psi^\dagger(\pp)$, giving statistical predictions
for the results of measurements of spin components by an ideal
apparatus which is not affected by the momentum of the particle.
The spin entropy is
\beq
S=-\tr(\tau\log\tau)=-\sum\lambda_j\log\lambda_j,
\eeq
where $\lambda_j$ are the eigenvalues of $\tau$.

As usual, ignoring some degrees of freedom leaves the others in a
mixed state. What is not obvious is that in the present case the
amount of mixing depends on the Lorentz frame used by the
observer.  Indeed consider another observer (Bob) who moves with a
constant velocity with respect to Alice who prepared state
(\ref{psi0}). In the Lorentz frame where Bob is at rest, the same
spin-\half\ particle has a state
\beq
\psi'(\pp)={a'_1(\pp)\choose a'_2(\pp)}.
\eeq
The transformation law is
\beq
a'(\pp)=[(\Lambda^{-1}p)_0/p_0]^{1/2}\,\sum_s
 D_{rs}[\Lambda,(\Lambda^{-1}p)]\,a_s(\Lambda^{-1}p) \label{tf},
 \eeq
where $D_{rs}$ is the Wigner rotation matrix for a Lorentz
transformation $\Lambda$ \cite{we:b, hal, bog}.

As an example, take a particle prepared by Alice with spin in the
$z$ direction, so that $a_2(\pp)=0$, and
\beq
a_1(\bp)=N\exp(-\bp^2/2\Delta^2),
\eeq
where $N$ is a normalization factor.
 Spin and momentum are not
entangled, and the spin entropy is zero. When that particle is
described in Bob's Lorentz frame, moving with velocity $\beta$ in
a direction at an angle $\theta$ with Alice's $z$-axis, a detailed
calculation shows that both $a'_1$ and $a'_2$ are nonzero, so that
the spin entropy is positive
\cite{pt03b,pst}. This phenomenon is illustrated in Fig.~1. It can
be shown
\cite{pt03b, tt} that a relevant parameter, apart from the angle
$\theta$, is in the leading order in momentum spread,
\beq
\Gamma=\frac{\Delta}{m}\,\frac{1-\sqrt{1-\beta^2}}{\beta},
\eeq
where $\Delta$ is the momentum spread in Alice's frame. {\it The
entropy has no invariant meaning\/}, because the reduced density
matrix $\tau$ has no covariant transformation law, except in the
limiting case of sharp momenta.  Only the complete density matrix
transforms covariantly.

\begin{figure}[htbp]
\epsfxsize=0.48\textwidth
\centerline{\epsffile{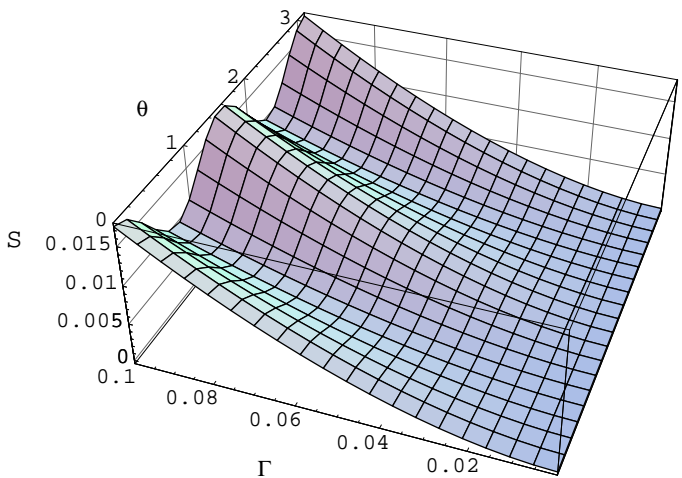}} \vspace*{-0.1cm} \caption{\small{Dependence
of the spin entropy $S$, in Bob's frame, on the values of the
angle $\theta$ and a parameter
$\Gamma=[1-(1-\beta^2)^{1/2}]\Delta/m\beta$.}}
\label{spinent}
\end{figure}

It is noteworthy that a similar situation arises for a classical
system whose state is given in any Lorentz frame by a Liouville
function \cite{bk:67}. Recall that a Liouville function expresses
our probabilistic description of a classical system --- what we
can predict before we perform an actual observation --- just as a
quantum state is a mathematical expression used for computing
probabilities of events.

Consider now a pair of orthogonal states that were prepared by
Alice. How well can moving Bob distinguish them? We shall use the
simplest criterion, namely the probability of error $P_E$, defined
as follows: an observer receives a single copy of one of the two
known states and performs any operation permitted by quantum
theory in order to decide which state was supplied. The
probability of a wrong answer for an optimal measurement is
\cite{fg99}
\beq
P_E(\rho_1,\rho_2)=\half-\mbox{$1\over4$}\,{\rm tr}
 \sqrt{(\rho_1-\rho_2)^2}.  \label{pe}
 \eeq
In Alice's frame $P_E=0$. It can be shown \cite{pt03b,tt} that in
Bob's frame, $P'_E\propto\Gamma^2$, where the proportionality
factor depends on the angle $\theta$ defined above.

An interesting problem is the relativistic nature of quantum
entanglement when there are several particles. For two particles,
an invariant definition of the entanglement of their spins would
be to compute it in  the Lorentz ``rest frame'' where
$\6\sum\pp\9=0$. However, this simple definition is not adequate
when there are more than two particles, because there appears a
problem of cluster decomposition: each subset of particles may
have a different rest frame. This is a difficult problem, still
awaiting for a solution. We shall mention only a few partial
results.

Alsing and Milburn \cite{am02} considered bipartite states with
well-defined momenta. They showed that while Lorentz
transformations change the appearance of the state in different
inertial frames and the spin directions are Wigner rotated, the
amount of entanglement remains intact. The  reason is that Lorentz
boosts do not create spin-momentum entanglement when acting on
eigenstates of momentum, and the transformations on the pair are
implemented on both particles as local unitary transformations
which are known to preserve the entanglement. The same conclusion
is also valid for photon pairs.

However, realistic situations involve wave packets. For example, a
general spin-$\half$  two-particle state  may be written as
%\begin{widetext}
\beq
\!|\Upsilon_{12}\9\!=\!\!
\sum_{\sigma_1,\sigma_2}\!\int\!
d\mu(p_1)d\mu(p_2)g(\sigma_1\sigma_2,\bp_1,\bp_2)|\bp_1,\sigma_1\9
\otimes|\bp_2,\sigma_2\9,
 \label{entstate}
\eeq
%\end{widetext}
where
\beq
d\mu(p)=\frac{d^3\bp}{(2\pi)^3 2p^0},
\eeq
is a Lorentz-invariant measure. For particles with well defined
momenta, $g$ is sharply peaked at some values $\bp_{10}$,
$\bp_{20}$. Again, a boost to any Lorentz frame $S'$ will result
in a unitary $U(\Lambda)\otimes U(\Lambda)$, acting on each
particle separately, thus preserving the entanglement. However, if
the momenta are not sharp, so that the spin-momentum entanglement
is frame dependent, then the spin-spin entanglement is
frame-dependent as well.

Gingrich and Adami \cite{ga02} investigated the reduced density
matrix for $|\Upsilon_{12}\9$ and made explicit calculations for
the case where $g$ is a Gaussian, as in \cite{pst}. They showed
that if two particles are maximally entangled in a common
(approximate) rest frame (Alice's frame), then the degree of
entanglement, as seen by a Lorentz-boosted Bob, decreases when the
boost parameter $\beta\to1$. Of course, the inverse transformation
from Bob to Alice will increase the entanglement. Thus, we see
that that spin-spin entanglement is not a Lorentz invariant
quantity, exactly as spin entropy is not a Lorentz scalar.

\section{Photons}\label{photons}

Relativistic effects that we describe in this section are
essentially different from those for massive particles that were
discussed above, because photons have only two linearly
independent polarization states. The properties that we discuss
are kinematical, not dynamical. At the statistical level, it is
not even necessary to involve quantum electrodynamics. Most
formulae can be derived by elementary classical methods
\cite{pt03a}. It is only when we consider individual photons, for
crypto\-graphic applications, that quantum theory becomes
essential. The diffraction effects mentioned above lead to
superselection rules which make it impossible to define a reduced
density matrix for polarization. As shown below, it is still
possible to have ``effective'' density matrices; however, the
latter depend not only on the preparation process, but also on the
method of detection that is used by the observer.

In applications to secure communication, the ideal scenario is
that isolated photons (one particle Fock states) are emitted. In a
more realistic setup, the transmission is by means of weak
coherent pulses containing on the average less than one photon
each.  A basis of the one-photon space is spanned by states of
definite momentum and helicity,
\beq
|\bk,\bep_\bk^\pm\9 \equiv |\bk\9\otimes|\bep_\bk^\pm\9,
 \label{basis}
 \eeq
where the momentum basis is normalized by $\6\bq|\bk\9=(2\pi)^3(2
k^0)\delta^{(3)}(\bq-\bk)$, and helicity states $|\bep_\bk^\pm\9$
are explicitly defined by Eq.~(\ref{helivectors}) below.

As we know, polarization is a {\it secondary variable\/}: states
that correspond to different momenta belong to distinct Hilbert
spaces and cannot be superposed (an expression such as
$|\bep_\bk^\pm\9+|\bep_\bq^\pm\9$ is meaningless if $\bk\neq\bq$).
The complete basis (\ref{basis}) does not violate this
superselection rule, owing to the othogonality of the momentum
basis. Therefore, a generic one-photon state is given by a wave
packet
\beq
|\Psi\9=\int d\mu(\bk)f(k)|\bk,\bal(\bk)\9.\label{photon}
\eeq
The Lorentz-invariant measure is $d\mu(k)=d^3\bk/(2\pi)^3 2k^0$,
and normalized states satisfy $\int d\mu(k)|f(\bk)|^2=1$.  The
generic polarization state $|\bal(\bk)\9$ corresponds to the
geometrical 3-vector
\beq
\bal(\bk)=\alpha_+(\bk)\bep^+_\bk+\alpha_-(\bk)\bep^-_\bk,
 \label{elliptic}
 \eeq
where $|\alpha_+|^2+|\alpha_-|^2=1$, and the explicit form of
$\bep^\pm_\bk$ is given below.

Lorentz transformations of quantum states are most easily computed
by referring to some standard momentum, which for photons is
$p^\nu=(1,0,0,1)$. Accordingly, standard right and left circular
polarization vectors are $\bep^\pm_p=(1,\pm i,0)/\sqrt{2}$. For
{\it linear\/} polarization, we take Eq.~(\ref{elliptic}) with
$\alpha_+=(\alpha_-)^*$, so that the 3-vectors $\bal(\bk)$ are
real. In general, complex $\bal(\bk)$ correspond to elliptic
polarization.

Under a Lorentz transformation $\Lambda$, these states become
$|\bk_\Lambda,\bal(\bk_\Lambda)\9$, where $\bk_\Lambda$ is the
spatial part of a four-vector $k_\Lambda=\Lambda k$, and the new
polarization vector can be obtained by an appropriate rotation
\cite{am02,hks:85}
\beq
\bal(\bk_\Lambda)=R(\hbk_\Lambda)R(\hbk)^{-1}\bal(\bk),
\eeq
where $\hbk$ is the unit vector in the direction of $\bk$.
Finally, for each $\bk$ a polarization basis is labeled by the
helicity vectors,
\beq
\bep^\pm_\bk=R(\hbk)\bep^\pm_p. \label{helivectors}
\eeq

Let us try to define a reduced density matrix in the usual way,
\beq
\rho=\int d\mu(\bk)|f(k)|^2|\bk,\bal(\bk)\9\6\bk,\bal(\bk)|?
\eeq
The superselection rule that was mentioned above does not forbid
this definition, because only terms with the same momentum $\bk$
are summed. However, since polarization is a secondary variable,
this object cannot have definite transformation properties under
boosts. This deficiency is familiar to us from the analysis of
reduced density matrices of massive particles. However,  for
massless particles, the situation is worse: POVMs that are given
by $2\times 2$ matrices represent  measurement devices and should
transform under a representation of the rotation group O(3). On
the other hand, even for the complete photon state,
 ordinary rotations of the reference frame correspond to  elements of E(2),
 so that probabilities would not be invariant under rotations.

 Therefore, let us find a more physical definition of a reduced density
matrix  for polarization \cite{pt03a}.  The labelling of
polarization states by Euclidean vectors $\be_\bk^n$ suggests the
use of a $3\times 3$ matrix with entries labelled $x$, $y$ and
$z$. Classically, they correspond to different directions of the
electric field. For example, a reduced density matrix $\rho_{x}$
would give the expectation values of operators representing the
polarization in the $x$ direction, seemingly irrespective of the
particle's momentum.

To have a momentum-independent polarization is to tacitly admit
longitudinal photons.   Unphysical concepts are often used in
intermediate steps in theoretical physics. Momentum-independent
polarization states thus consist of physical (transversal) and
unphysical (longitudinal) parts, the latter corresponding to a
polarization vector $\bep^\ell=\hbk$. For example, a generalized
polarization state along the $x$-axis is
\beq
|\hbx\9=x_+(\bk)|\bep^+_\bk\9+x_-(\bk)|\bep^-_\bk\9+
x_\ell(\bk)|\bep^\ell_\bk\9,\label{decomp}
\eeq
where $x_\pm(\bk)=\bep^\pm_\bk\cdot\hbx$, and $x_\ell(\bk)=
\hbx\cdot\hbk=\sin\theta\cos\phi$. It follows that
$|x_+|^2+|x_-|^2+|x_\ell|^2=1$, and we thus define
\beq
\be_x(\bk)=\frac{x_+(\bk)\bep^+_\bk+x_-(\bk)\bep^-_\bk}
{\sqrt{x_+^2+x_-^2}} \label{physdir},
\eeq
as the polarization vector associated with the $x$ direction. It
follows from (\ref{decomp}) that $\6\hbx|\hbx\9=1$ and
$\6\hbx|\hby\9=\hbx\cdot\hby=0$, and likewise for other
directions, so that
\beq
|\hbx\9\6\hbx|+|\hby\9\6\hby|+|\hbz\9\6\hbz|=\1_p,\label{xyz}
\eeq
where $\1_p$ is the unit operator in momentum space.

 We can now
define an ``effective" reduced density matrix adapted to a
specific method of measuring polarization, as follows
\cite{pt03a}. To the direction $\hbx$ corresponds a projection
operator
\beq
P_{x}=|\hbx\9\6\hbx|\otimes \1_p=|\hbx\9\6\hbx|\otimes \int
d\mu(k)|\bk\9\6\bk|,
\eeq
 The action of
$P_{x}$ on $|\Psi\9$ follows from Eq.~(\ref{decomp}) and
$\6\bep^\pm_\bk|\bep^\ell_\bk\9=0$. Only the transversal part of
$|\hbx\9$ appears in the expectation value:
\beq
\6\Psi|P_{x}|\Psi\9\!=\!\int \!d\mu(k)|f(\bk)|^2|x_
 +(\bk)\alpha_+^*(\bk)+x_-(\bk)\alpha_-^*(\bk)|^2.
 \eeq
It is convenient to write the transversal part of $|\hbx\9$ as
\bea
|\bb_x(\bk)\9  \equiv
 (|\bep^+_\bk\9\6\bep^+_\bk|+|\bep^-_\bk\9\6\bep^-_\bk|)|\hbx\9
 \nonumber \\
 = x_+(\bk)|\bep^+_\bk\9+x_-(\bk)|\bep^-_\bk\9.
\label{vector}
\eea
%\end{eqnarray}
Likewise define  $|\bb_y(\bk)\9$ and $|\bb_z(\bk)\9$. These three
state vectors are neither of unit length nor mutually orthogonal.
For $\bk= (\sin\theta\cos\phi,\sin\theta\sin\phi,\cos\theta)$ we
have
%\begin{widetext}
\beq
|\bb_x(\bk)\9  =  %[(\cos\theta\cos\phi+i\sin\phi)|\bep^+_\bk\9+
 %(\cos\theta\cos\phi-i\sin\phi)|\bep^-_\bk\9]/\sqrt{2}=
   c(\theta,\phi)|\bk,\be_x(\bk)\9,
\eeq
%\end{widetext}
where $\be_x(\bk)$ is given by Eq.~(\ref{physdir}),  and
$c(\theta,\phi)=\sqrt{x_+^2+x_-^2}$.

Finally, a POVM element $E_{x}$ which is the physical part of
$P_{x}$, namely is equivalent to $P_{x}$ for physical states
(without longitudinal photons) is
\beq
E_{x}=\int d\mu(k)|\bk,\bb_x(\bk)\9\6\bk,\bb_x(\bk)|,
\eeq
and likewise for other directions. The operators $E_{x}$, $E_{y}$
and $E_{z}$ indeed form a POVM in the space of physical states,
owing to Eq.~(\ref{xyz}). The above derivation was, admittedly, a
rather circuitous route for obtaining a POVM for polarization.
This is due to the fact that the latter is a secondary variable,
subject to super\-selection rules. Unfortunately, this is the
generic situation.

To complete the construction of the density matrix, we introduce
additional directions. Similarly to a standard practice of channel
matrices reconstruction
\cite{cn:97}, we consider $E_{x+z,x+z}$, $E_{x-iz,x-iz}$ and similar
combinations. For example,
\beq
E_{x+z,x+z}=\half(|\hbx\9+|\hbz\9)(\6\hbx|+\6\hbz|)\otimes\1_p.
\eeq

Let us denote $|\hbx\9\6\hbz|\otimes\1_p$ as $E_{xz}$, even though
this is not a positive operator. We then get a simple expression
for the reduced density matrix corresponding to the polarization
state $|\bal(\bk)\9$:
\bea
 \rho_{mn}=\6\Psi|E_{mn}|\Psi\9=  \nonumber \\
 \int
d\mu(k)|f(\bk)|^2\6\bal(\bk)|\bb_m(\bk)\9\6\bb_n(\bk)|\bal(\bk)\9
, \label{reduced}
\eea
where $ m,n,=x,y,z$. It is interesting to note that this
derivation gives a direct physical meaning to the naive definition
of a reduced density matrix,
\beq
\rho^{\rm naive}_{mn}=\int d\mu(\bk)|f(k)|^2\bal_m(\bk)\bal_n^*(\bk)
=\rho_{mn}
\eeq

Our basis states $|\bk,\bep_\bk\9$ are direct products of momentum
and polarization. Owing to the transversality requirement
$\bep_\bk\cdot\bk=0$, they remain direct products under Lorentz
transformations. All the other states have their polarization and
momentum degrees of freedom entangled. As a result, if one is
restricted to polarization measurements as described by the above
POVM, {\it there do not exist two orthogonal polarization
states\/}. An immediate corollary is that photon polarization
states cannot be cloned perfectly, because the no-cloning theorem
\cite{nc} forbids an exact copying of unknown non-orthogonal
states. In general, any measurement procedure with finite momentum
sensitivity will lead to the errors in identification. First we
present some general considerations and then illustrate them with
a simple example.

Let us take the $z$-axis to coincide with the average direction of
propagation so that the mean photon momentum is $k_A\hbz$.
Typically, the spread in momentum is small, but not necessarily
equal in all directions. Usually the intensity profile of laser
beams has cylindrical symmetry, and we may assume that
$\Delta_x\sim\Delta_y\sim\Delta_r$ where the index $r$ means
radial. We may also assume that $\Delta_r{\gg}\Delta_z$. We then
have
\beq
f(\bk)\propto f_1[(k_z-k_A)/\Delta_z]\,f_2(k_r/\Delta_r).
\eeq
 We  approximate
\beq
\theta\approx\tan\theta\equiv k_r/k_z\approx k_r/k_A.
\label{theta}
\eeq
In  pictorial language, polarization planes for different momenta
are tilted by angles up to $\sim\Delta_r/k_A$, so that we expect
an error probability of the order $\Delta_r^2/k_A^2$. In the
density matrix $\rho_{mn}$ all the elements of the form
$\rho_{mz}$ should vanish when $\Delta_r\to0$. Moreover, if
$\Delta_z\to0$, the non-vanishing $xy$ block goes to the usual
(monochromatic) polarization density matrix.
%\beq
%\rho_{{\rm pure}}=\left(\bay{ccc}
%|\alpha|^2 & \beta & 0\\
%\beta^* & 1-|\alpha|^2 &0\\
%0 & 0 &0
%\eay\right).
%\eeq

As an example we consider two states which, if the momentum spread
could be ignored, would be $|k_A\hat{\bf z},\bep^\pm_{k_A\hat{\bf
z}}\9$. To simplify the calculations we assume a Gaussian
distribution:
\beq
f(\bk)=Ne^{-(k_z-k_A)^2/2\Delta_z^2}e^{-k_r^2/2\Delta_r^2},
\eeq
where $N$ is a normalization factor and $\Delta_z\ll\Delta_r$.
Moreover, we take the polarization components to be
$\bep^\pm_\bk\equiv R(\hbk)\bep^\pm_p$. That means we have to
analyze  the states
\beq
|\Psi_\pm\9=\int d\mu(\bk) f(\bk)|\bep^\pm_\bk,\bk\9\label{st},
\eeq
where $f(\bk)$ is given above.

It can be shown \cite{pt03a} that in the leading order
\beq
P_E(\rho_+,\rho_-)=\Delta_r^2/4k_A^2.
\eeq

Now we turn to the distinguishability problem from the point of
view of a moving observer, Bob.  The probability of an error by
Bob is still given by Eq.~(\ref{pe}). The distinguishability of
polarization density matrices depends on the observer's motion. We
again assume that Bob moves along the $z$-axis with a velocity
$v$.   Detailed calculations \cite{pt03a} show that
\beq
P'_E=\frac{1+v}{1-v}P_E, \label{dope}
\eeq
which may be either larger or smaller than $P_E$. As expected, we
obtain for one-photon states the same Doppler effect as in the
 classical calculations \cite{pt03a}.

Since maximally entangled states are pure, the fact that all
polarization density matrices are mixed implies that maximal
EPR-type correlations shall never be observed, and that maximal
attainable value will depend on the momentum spread of the states.

\section{Communication channels}\label {cocha}

Although reduced polarization density matrices have no general
transformation rule, the above results show that such rules can be
established for particular classes of experimental procedures. We
can then ask how these effective transformation rules,
$\tau'=T(\tau)$, fit into the framework of general state
transformations. Equation~(\ref{chanex}) gives an example of such
a transformation.
\beq
\rho'=\rho(1-\frac{\Gamma^2}{4})+(\sigma_x\rho\sigma_x+\sigma_y\rho\sigma_y)
\frac{\Gamma^2}{8}. \label{chanex}
\eeq
It relates the reduced density matrix $\rho$, obtained from
Alice's Eq.~(\ref{psi0}), to Bob's density matrix $\rho'$. This
particular transformation is completely positive; however, it is
not so in general.

It can be proved that distinguishability, as expressed by natural
measures like $P_E$, cannot be improved by any completely positive
transformation
\cite{fg99}. It is also known that the complete positivity
requirement may fail if there is a prior entanglement with another
system
\cite{cpnot}. Since in the two
previous sections we have seen that distinguishability {\it can\/}
be improved, we conclude that these transformations are {\it not
completely positive\/}. The reason is that the Lorentz
transformation acts not only on the ``interesting'' discrete
variables, but also on the ``hidden'' momentum variables that we
elected to ignore and to trace out, and its action on the
interesting degrees of freedom depends on the hidden ones.

This technicality has one important consequence. The notion of a
{\em channel} is fundamental both in classical  and quantum
communication theory.  Quantum channels  are described as
completely positive maps that act on qubit states
\cite{chan}.  Qubits themselves are realized as particles' discrete degrees
of freedom. If relativistic motion is important, then not only
does the vacuum behave as a noisy quantum channel, but the very
representation of a channel by a completely positive map fails
\cite{pt03b}.

\end{document}